\documentclass[pre]{revtex4}
\usepackage{graphicx}
\usepackage{dcolumn}

\begin{document}

\title{Magnetic field induced band depopulation in intrinsic InSb: A revisit\footnote{\it To appear in Journal of Physics: Condensed Matter}}
\author{Bhavtosh Bansal\footnote{Present address: Department of Condensed Matter Physics and Materials Science, Tata institute of Fundamental Research, 1 Homi Bhabha Road, Mumbai 400005, India. \\Electronic mail: bhavtosh@tifr.res.in}, V. Venkataraman\footnote{Electronic mail:venki@physics.iisc.ernet.in}}
\affiliation{Department of Physics, Indian Institute of Science,
Bangalore 560012, India.}
\date{\today}
\begin{abstract}
The effect of Landau level formation on the population of intrinsic electrons in InSb is probed near room temperature in magnetic fields upto 16 Tesla. Although the measured magnetic field dependence of the Hall coefficient is qualitatively similar
to published results, it is shown that the data may also be explained by simply including ambipolar conduction. Thus the inference on band depopulation drawn from previous measurements on InSb is inconclusive unless both the Hall and the magnetoresistive components of the resistivity tensor are simultaneously measured and modelled. When the model includes both depopulation and ambipolar conduction, a reasonable agreement with theory can be established. 
\end{abstract}
%\pacs{}
\maketitle
\section{Introduction}
Large magnetic fields significantly change the electronic density of states through formation of Landau levels(LL). In three dimensions, the level structure consists of a shift in the band edge by the zero point cyclotron energy($1/2\hbar\omega_c$) and subsequent magnetic sub-bands separated by cyclotron energy($\hbar\omega_c$), with a weak (square root) singularity in the density of states at the bottom of each sub-band. The quantization energy is about one Kelvin per Tesla for electrons in vacuum and in most solids, but can be around a hundred times larger in narrow gap semiconductors like InSb. From the early 1950s, experiments\cite{optical} on InSb have clearly measured a blue shift in magneto-absorption, even at room temperature under moderate magnetic fields ($\sim$5 Tesla). And over the years there has been a large volume of literature devoted to spectroscopic investigations of this change in different semiconductors\cite{landau level spectroscopy}.\\

While most quantum magnetotransport experiments are performed at low temperatures under the condition of strong degeneracy of the electron gas and probe the Fermi surface through 
oscillatory behaviour of the transport coefficients, it is interesting to ask whether the  same  rearrangement in the density of states can also be seen near room temperature, in the intrinsic regime (weak degeneracy) through a conventional transport measurement. For  
instance, this should be observable by measuring the change in the population of  
thermally excited carriers across the energy gap as a function of the magnetic  
field\cite{beer_t2,  love, aronzon}.  Naively, we may expect\cite{arora, girist} the  
intrinsic carrier density to exponentially decrease with this increased
effective gap, $n(B) \sim n(0)\hspace{.1cm}exp[{-\hbar\omega_c/k_BT}]$. But this argument ignores the significance of singularities in the density of states accompanying Landau quantization, the effect of Zeeman splitting and sample degeneracy.

There are at least two old experiments on InSb\cite{love, aronzon} where a change in the Hall coefficient, as a function of magnetic field, was measured and attributed to band depopulation. We shall show below that ambipolar conduction, due to an (almost) equal number of holes, also changes the Hall coefficient in a qualitatively similar fashion. Hence the analysis in references \cite{love, aronzon} is incomplete. Thus in InSb, the most studied narrow  gap semiconductor, this band depopulation has so far not
been satisfactorily observed. Similar transport studies also exist on zero gap
Hg$_{1-x}$Cd$_x$Te, where a gap opens up in the presence of a quantizing magnetic field. A comparatively recent paper\cite{thio} on giant magnetoresistance does account for the classical two carrier effects but treats the quantum mechanical freeze-out only phenomenologically. The work of Byszewski, et. al. \citep{walukiewich} is perhaps the most comprehensive and, although it is on zero gap HgTe and differs in the method of  analysis, is closest in spirit to the results reported here. \\

In this paper, we shall describe measurements of the diagonal and the off-diagonal  
components of the resistivity tensor in intrinsic n-type InSb, measured as a function of magnetic field upto 16 Tesla. These data are then simultaneously fitted to a model of the resistivity tensor which includes ambipolar conduction and Landau level effects. The intrinsic population change is computed within the four band (with spin) k$\cdot$p theory with Zeeman split LLs using exact carrier statistics. The calculations are further extended for the magnetic field dependence of the carrier density assuming different values of the electron g-factors. Parenthetically we note that the phenomenon studied here is very different from the better known\cite{seeger} magnetic freeze-out of {\em extrinsic} carriers (forming an impurity band) at very low temperatures. 

\section{Experimental Details}
\begin{figure}[!h]
\begin{center}
\resizebox{!}{6.6cm} {\includegraphics{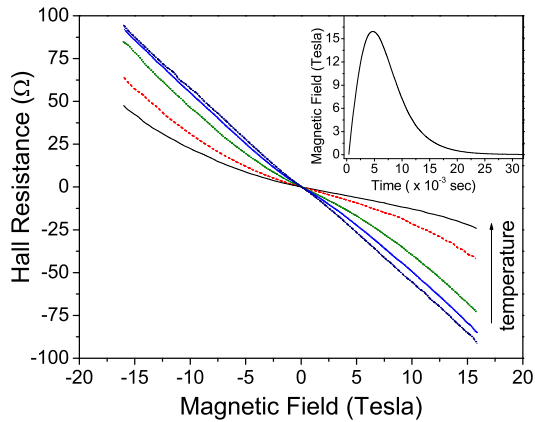}}
\caption{\label{fig:plus_minus}{\it The measured Hall resistance in van der Pauw geometry. Data shown for temperatures between 170 K and 250 K in steps of 20 K. A slight $\pm B$ asymmetry indicates magnetoresistance. Magnetoresistance is maximum at 250K due to the large number of intrinsic carriers.}}
\end{center}
\end{figure}

\begin{figure}[!t]
\begin{center}
\resizebox{!}{8cm} {\includegraphics{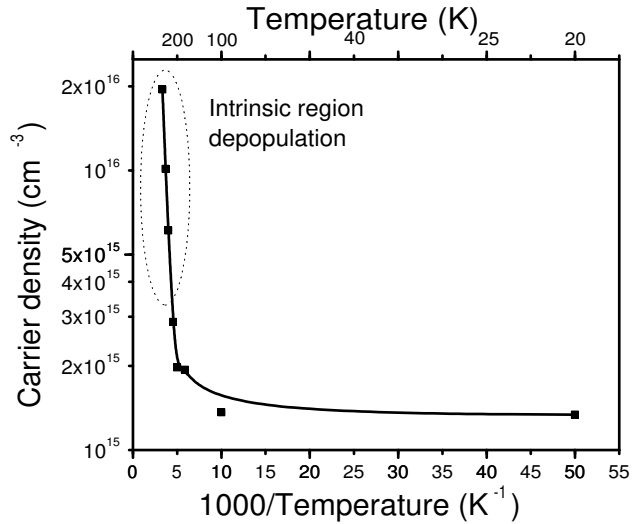}}
\caption{\label{fig:insb_hall}{\it Temperature dependence of the
carrier density for the InSb sample. The sample is clearly
intrinsic above 220K}}
\end{center}
\end{figure}

Experiments were performed on n-type InSb crystals of moderate purity. The measured carrier density as a function of temperature is plotted in figure \ref{fig:insb_hall}. The temperature range of interest is also marked in the figure. Magnetic field dependent
magnetoresistance and Hall coefficients were measured in magnetic fields up to 16 Tesla using a homemade pulsed magnet. The pulsed magnet has a 60 mF, 450 V capacitor bank made from commonly used electrolytic capacitors. These are connected to a 200 turn
araldite reinforced coil via a thyristor and a crowbar diode. The bore of the coil is 12 mm and the length 60 mm. The sample is kept inside a stainless steel cryostat with a glass tail and the sample holder is made of silicon. While operating, the magnet is immersed in liquid nitrogen to reduce the coil resistance. The cryostat temperature can be controlled between 77K and 350K. The pulse rise time is around 5 ms and decays by around 30 ms. The data is recorded on a 16 bit Keithley high speed voltmeter at 100 kHz through a low noise preamplifier. A typical magnetic field profile as a function of time is shown in figure \ref{fig:plus_minus}(inset). \\
The samples were made in van der Pauw configuration with small peripheral contacts to minimize electrostatic shorting of the Hall voltage. By Onsager's reciprocity relations, the off-diagonal part of the conductivity tensor is antisymmetric and the diagonal part is symmetric in a magnetic field. Thus the physical magnetoresistance caused by the misalignment voltage and the Hall resistivity can be separated by adding and subtracting the signal for opposite magnetic field polarities.  Both components of the tensor, $r_{xx}$ and $r_{xy}$, can be extracted from any single curve in figure \ref{fig:plus_minus}. Of course, the symmetric part of the curve needs to be appropriately scaled by the value of the zero field resistivity of the sample at that temperature. This method for extraction of the two components of the magnetoresistivity tensor has been previously experimentally demonstrated to yield very satisfactory results \cite{iee-vanderpauw}.

Prior to the high magnetic field measurements, the carrier concentration and  mobility of  the InSb sample were determined by low field Hall measurements between 10K and 300K. The sample had a background doping of $3\times 10^{15}cm^{-3}$ and was seen to be
clearly intrinsic above 220K (figure \ref{fig:insb_hall}). In the temperature range of 
interest the mobility was around 10 $m^2/Vs$. At 250K and in 10 Tesla field, $\mu B\sim 100$ and $\hbar \omega_c\sim 10 k_BT$. Hence the extreme quantum limit was safely
reached and Landau level broadening could be ignored. Zeeman level broadening is also  
expected to be of the order of transport relaxation time since even non-magnetic  
potentials can cause electron spins to flip in InSb. The contributions of the Hall and
magnetoresistive factors of conduction electrons to the field dependence of $r_{xy}$ and $r_{xx}$ also saturate to unity by 1 Tesla. These factors arise when the relaxation time is averaged over the carrier distribution. 
\section{Theory}
In presence of a strong magnetic field the energy dispersion in the plane perpendicular to the field is modified due to the Landau quantization of cyclotron orbits. Since the spin orbit splitting energy in InSb is almost four times the energy gap, a four band (including spins) k$\cdot$p model is a very good
description for the conduction electron states in InSb. The strong valence and conduction band coupling results in band non-parabolicity which makes the LL separation unequal. The effective mass and g-factor have energy dependent values. The
energy dispersion in a magnetic field taking this and the large spin splitting into account is \cite{analogy} 
\begin{equation}\label{dispersion}
E=\lbrace({E_g\over 2})^2+E_g[({\hbar^2k_z^2\over 2m^{\ast}})+(n+{1\over 2}\pm{g\over  
4})\hbar \omega_c]\rbrace^{1/2}
\end{equation}
where $g$ is the effective g-factor defined such that the Zeeman splitting is in the units of cyclotron frequency. Thus, for InSb, whose g-factor, $g^{\ast}=51$ and effective mass is $m^{\ast}=0.014m_0$ then the effective g-factor, $g=g^{\ast} m^{\ast}/m_0$,  is 0.7. $\mu_0=e \hbar/ 2m_0 $ is the Bohr magneton. From equation
\ref{dispersion}, the density of states(DOS) in a magnetic field
can be written as 
\begin{eqnarray} \label{dos}
D(E)=\pi({2m^{\ast}\over h^2})^{3/2}({2E\over E_g} + 1)\hbar
\omega_c  \nonumber \hspace{3.5cm}\\
\times\sum_{n=0,\pm}^{n_{max}} [{(E^2/ E_g)+(E\pm{1\over2} \mu_B
|g|B)}-\hbar \omega_c(n+{1\over2})]^{-{1/2}}
\end{eqnarray}
The summation has to be carried over both the LL index $n$ and the
spin up and down states denoted by $\pm$.
The heavy-hole and light-hole bands of InSb are degenerate at the $\Gamma$ point. Because of the much larger density of states, the heavy-hole band contributes most of the holes. Thus the light-hole band's contributions to the total hole population can be neglected. Secondly, the large heavy-hole mass ($m_{hh} > 10 m^{\ast}$) makes the effect of the magnetic field in quantizing these levels also relatively unimportant. The smaller heavy-hole mobility also implies a greater level broadening, further diminishing the quantization effects. Therefore holes can be treated like free particles with a three-dimensional parabolic dispersion even in the presence of  
a reasonably strong magnetic field.

The carrier density in a magnetic field can be determined for a given Fermi energy in the  usual way by integrating the Fermi-Dirac distribution function weighted by the now  modified DOS given by equation \ref{dos}. The resulting expression involves an infinite  summation over generalized Fermi integrals. The equilibrium electron and hole populations  at a given magnetic field and temperature are determined iteratively, under the  constraint that the intrinsic electron and hole populations are equal.

\section{Results and Analysis}
\begin{figure}[h]
\begin{center}
\resizebox{!}{10cm} {\includegraphics{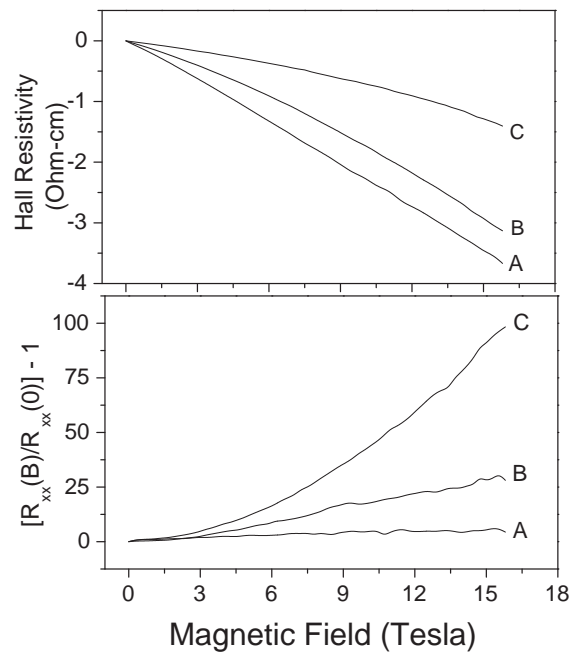}}
\caption{\label{fig:magnetoresistance_hall}{\it The measured Hall resistity and the magnetoresistance at three representative temperatures. A.170K,  B.210K and C.250K.}}
\end{center}
\end{figure}

Figure \ref{fig:plus_minus} shows the four terminal Hall resistance measured in van der Pauw geometry at temperatures between 170K and 250K. The $\pm B$ asymmetry in the signal is due to superimposed magnetoresistance. The magnetic field dependence of the Hall resistivity and magnetoresistance at different temperatures is shown in figure \ref{fig:magnetoresistance_hall}.  At 170K, the sample is extrinsic as can be seen from figure \ref{fig:insb_hall}. Therefore the Hall resistance is approximately linear up to 16 Tesla and the magnetoresistance is small (figure \ref{fig:magnetoresistance_hall}, curves `A'). It is emphasized that accurately measuring the {\em physical} magnetoresistance in high mobility samples is non-trivial. Due to the high mobility of the electrons and the extremely high magnetic field employed in the measurement, any small inhomogeniety or the effects of finite sized contacts  can cause partial shorting of the Hall voltage. This can yield large magnetoresistance. To get an estimate of the possible magnitude of the geometric magnetoresistance, we note that at 170K, the electron mobility in our sample was around $1.5\times 10^5 cm^2/Vs$. This implies, that the Hall angle at 15 Tesla was 89.7$^o$ and if the Hall field was to be completely shorted, the magnetoresistance, $\mu^2B^2$ would have been greater than 50,000. Even $0.1\%$ of this value would imply relatively large magnetoresistance and our observation of finite $\Delta R/ R < 10$ at 170K is most probably of geometric origin. 

As the temperature is increased from 170K to finally 250K, we observe in figure \ref{fig:magnetoresistance_hall} that the average slope and magnitude of the Hall resisitivity decreases. While this is a natural consequence of the increase in the intrinsic carrier density, with increasing temperature we also observe that (1) the Hall resistivity at high magnetic field becomes progressively non-linear (2)The magnetoresistance increases in magnitude.  Figure \ref{fig:magnetoresistance_hall} shows that the two effects are intimately linked and get more pronounced as the samples becomes more intrinsic. 

In the past, the superlinearity of the Hall reistivity has been used as the signature of intrinsic carrier freeze-out\cite{beer_t2, love, aronzon}. Under this regime of intrisic carrier depopulation, one may also expect to observe a large magnetoresistance\cite{arora}, simply because of the decrease in the number of free carriers contributing to conductivity. Therefore the above trends may be taken to provide a consistent picture of depopulation. 

Unfortunately, a completely different mechanism linked with ambipolar conduction is also independently responsible for both the apparent increase in the Hall resistance with magnetic field, and the large positive magnetoresistance. Furthermore, since  band depopulation can only be observed in intrinsic samples, it neccessarily implies that both electrons and holes would simultaneously contribute to the measured transport coefficients. 
\begin{figure}[!h]
\begin{center}
\resizebox{!}{8cm} {\includegraphics{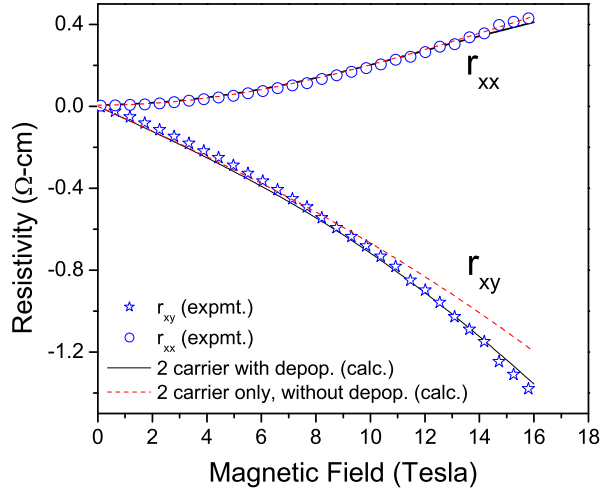}}
\caption{{\label{fig:calc_exp}}\it Diagonal ($r_{xx}(B)$) and off-diagonal ($r_{xy}(B)$) resistivity tensors at  250K deduced from figure \ref{fig:plus_minus}. Theoretical curves derived from  calculations with ambipolar conduction alone and with both depopulation and ambipolar 
conduction are also shown.}
\end{center}
\end{figure}

In figure \ref{fig:calc_exp}, we attempt a quantitative analysis of the the magnetic field dependent Hall and transverse magnetoresistivity, $r_{xx}(B)$and $r_{xy}(B)$ at 250 K when the departure from linearity is maximum and the magnetoresistance is the largest. The analysis has to simultaneously account for both the classical (ambipolar conduction) and the quantum mechanical (band depopulation) effects. Under conditions of ambipolar transport, the diagonal and the off-diagonal components, $s_{xx}$ and $s_{xy}$ of the magnetoconductivity tensor are just the sum of individual contributions to the total conduction from electrons and holes. 
\begin{eqnarray}\label{hall}
s_{xx}(B)=\sum_{i=e,h} {s_i(0)\over 1+ \mu_i^2B^2}\;\;\nonumber \\
s_{xy}(B)=\sum_{i=e,h} {s_i(0)\mu_iB\over 1+ \mu_i^2B^2}
\end{eqnarray}
where $s_e$, $s_h$ and $\mu_e$, $\mu_h$ are the zero magnetic field conductivities and mobilities of electrons and holes respectively; ${\bf r}(B)=[{\bf s}(B)]^{-1}$. Interestingly we observed that if only the Hall resistivity was fitted, then its complete variation could be explained using ambipolar conduction alone. The same is true for transverse magnetoresistance. And as has been already mentioned, previous studies have been also able to explain the change in the Hall coefficient on the basis of only band depopulation. On the other hand, a simultaneous fit to both the tensors for the same set of fit parameters ensures a much more constrained analysis of the data. We found that it was not possible to explain both the tensors using this two-carrier model alone.

Therefore, a best fit to the two tensors (using equation \ref{hall})  
was attempted with a magnetic field dependent intrinsic carrier density.  The
carrier density was calculated using the prescription described in the previous section.  Figure \ref{fig:calc_exp} shows the  fit obtained for this model. The best fit has the following set of parameter values: intrinsic electron and hole densities equal to $5.3\times 10^{16} cm^{-3}$, a background electron
doping density equal to $5.0\times 10^{16} cm^{-3}$ and the electron and hole mobilities equal to $700 cm^{2}/Vs$ and $10\times 10^4 cm^{2}/Vs$ respectively. The values of the g-factor (g${^\ast}$) and the electron effective mass were
assumed to be 50 (i.e. $g=0.7$) and  $m^{\ast}=0.014m_0$ respectively. These numbers compare well with the expected values of the electron and hole mobilities, intrinsic carrier densities and the background doping which was independently measured in a
low field Hall measurement at 12K. As expected, data below 10 Tesla can be explained with either model since the band depopulation starts to pick up only after 10 Tesla.

\begin{figure}[!h]
\begin{center}
\resizebox{!}{7cm} 
{\includegraphics{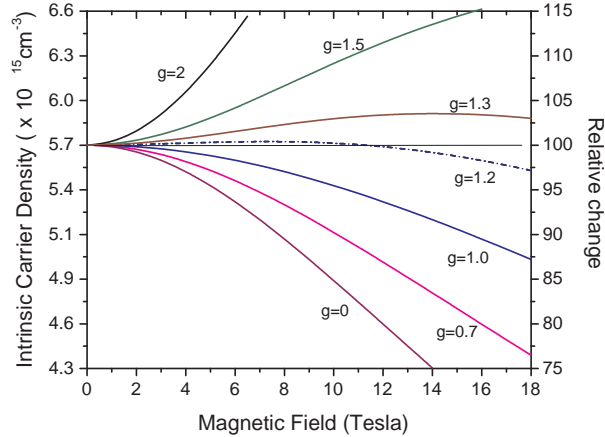}}
\caption{\label{fig:gfact}{\it Variation of the intrinsic carrier
density as a function of magnetic field for different values of
the effective g-factor. Calculations are done for InSb at 250K.
In these units, the effective g-factor ($g=g^{\ast} (m^{\ast}/m_0)$) of bulk pure InSb is around 0.7 near room temperature.}}
\end{center}
\end{figure}

Finally, figure \ref{fig:gfact} shows the calculated variation of the intrinsic carrier density with magnetic field in, for definiteness, InSb at 250K in the parameter space of effective g-factor, $g=g^{\ast} (m^{\ast}/m_0)$, values. Here $m^{\ast}$ is the carrier effective mass and $m_0$ is the free electron mass. There is a critical value of the g-factor, around 1.2, above which the carrier density {\it increases} with the applied magnetic field (up to some field value) even though the effective
gap also increases. This is a consequence of the collapse of some higher lying free electron levels into the highly degenerate LLs. Although the effective g-factors for most of the common narrow gap semiconductors are around 0.7 or less, anomalously large
g-factors have been observed when these are doped with magnetic impurities. This is due to an additional exchange interaction between the extended band  states and the localized magnetic moment of Mn$^{2+}$ ions \cite{winkler}.

At $g=2$, one Zeeman split LL gets pinned to the conduction band edge and the intrinsic electron density always increases in the magnetic field. Physically, this also
corresponds to free relativistic electrons in vacuum at temperatures of the order of the  
pair creation energy
($10^{10}K$). In such a situation, the equilibrium density of
electron-positron pairs will {\it monotonically} increase as a
function of the magnetic field. This conclusion is drawn using
the well known mapping of the two-band k$\cdot$p
model\cite{analogy} to the Dirac equation. Although very high fields($10^{10}$ Tesla in  
pulsars) and high temperatures (larger than pair creation energy, $10^6 eV$ in
supernova explosions) are known to separately exist, the only
astrophysical situation where both these extremes exist simultaneously seems to be the
early universe\cite{early_universe}. It is amusing to note that this extreme situation  
can easily be accomplished in a semiconductor, that is, if one chooses to take this  
analogy seriously.

\section{Conclusions}
The study attempted to experimentally detect the effect of Landau level formation on the density of states in a weakly degenerate system through a near room temperature transport experiment. While the fact that, as a consequence of Landau level formation, the intrinsic carrier density in InSb must decrease with increasing magnetic field is well accepted, it was shown that the analysis of the two previous experiments claiming to observe this effect in InSb is flawed. It was seen that most of the magnetic field  
dependence in the transverse magnetoresistivity and the Hall resistivity tensors could be accounted for by including the ambipolar conduction effects (which are inevitable in an intrinsic sample) alone.  Since the effect is weak, the calculations had to account for band non-parabolicity, spin splitting and exact electron statistics along with ambipolar conduction to reproduce the experimental data. It was also shown that if the g-factor values are assumed to be slightly larger, the intrinsic carrier density can surprisingly increase as a function of magnetic field even when the effective gap is larger than its zero field value. 
\section{Acknowledgements}
We thank V.K. Dixit and H.L. Bhat for the InSb samples used in this study. We also thank Department of Science and Technology (DST) and Defence Research and Development Organization (DRDO), Government of India for financial assistance. 

%\end{references}
\end{document}